# InSe: a two-dimensional semiconductor with superior flexibility

Qinghua Zhao,[a,b,c] Riccardo Frisenda, *[c] Tao Wang *[a,b] and Andres Castellanos-Gomez*[c]

Two-dimensional Indium Selenide (InSe) has attracted extensive attention recently due to its record-high charge carrier mobility and photoresponsivity in the fields of electronics and optoelectronics. Nevertheless, the mechanical properties of this material in the ultra-thin regime have not been investigated yet. Here, we present our efforts to determine the Young's modulus of thin InSe (~1-2 layers to ~40 layers) flakes experimentally by using buckling-based methodology. We find that the Young's modulus has a value of 23.1 $\pm$ 5.2 GPa, one of the lowest values reported up to date for crystalline two-dimensional materials. This superior flexibility can be very attractive for different applications, such as strain engineering and flexible electronics.

## Introduction

Two-dimensional (2D) semiconductors have attracted the interest of a great deal of the scientific communities due to their interesting electrical and optical properties that make them promising candidates for future electronic and optoelectronic applications.[1-3] For example, 2D transition metal dichalcogenides (TMDCs) have demonstrated remarkable performances in field-effect transistors (FETs), photodetectors, solar cells, and logic circuits.[4] Another interesting feature of 2D semiconductors is that they can sustain very large deformations without breaking. This has motivated the application of 2D semiconductors in flexible electronic devices and strain engineering studies.[9-12] The flexible electronic devices based on 2D dimensional semiconductors stand out due to their high carrier transport mobility, high specific surface area, high optical transparency, excellent mechanical resilience, and environmental stability.[13] Also, the strain engineering on 2D semiconductors, explored first theoretically and then experimentally,[16, 17] allows one for the modification of the band gap of these materials by means of mechanical deformations of their lattices.[18-20] These desirable features indicate that 2D semiconductors are a very promising alternative to traditional flexible organic semiconducting devices. Nevertheless, most of the 2D semiconductors studied so far are extremely stiff, having Young's modulus in the range of ~100-300 GPa (a value much larger than that of organic semiconducting materials (0.1 – 10 GPa),[23] see Table 1 and Fig. 3a). This means that a high stress is required to deform them. Therefore, flexible (or compliant) 2D semiconductors (characterized by a low Young's modulus) would be desirable for certain applications including 1) flexible electronic applications, where a semiconductor material with low Young's modulus mismatch with the polymeric/foil substrate is required,[25] 2) force sensing, 3) strain engineering, in which one wants to avoid slippage during strain cycles,[27, 28] and 4) nano-electro-mechanical systems, where one wants the resonators behave in the membrane-like regime.[30, 31]

Here, we use the buckling-based metrology method to measure the Young's modulus of thin InSe,[24] a promising 2D semiconductor that shows record-high electron mobility and ultrahigh photoresponse.[33-35] We determine the Young's modulus value *E* = 23.1 $\pm$ 5.2 GPa and we compare this value with the reported values for the other 2D semiconductors, finding that thin InSe is one order of magnitude more flexible than TMDCs. Our results are relevant for the application of InSe in future flexible electronics, straintronic devices, and nanomechanical systems.

## Results and discussions

2D thin InSe flakes were isolated by mechanical exfoliation from high quality single crystalline InSe ingots grown by Bridgman method.[37] Bulk InSe have been characterized by scanning electron microscopy (SEM), X-ray diffraction (XRD) and transmission electronic microscopy (TEM) to ensure its high crystallinity and to identify its polytope. More details about the characterization can be found in Section 1 of the Supplementary Information. We address the reader to the Materials and Method section for details about the InSe exfoliation. In order to measure the mechanical properties of InSe flakes through the buckling metrology method, they have to be deposited onto a very complaint substrate and subjected to uniaxial compression. In the case of InSe deposited on top of Gel-Film (by Gel-Pak®), if the compression strain in the InSe is above a critical value of approximately 0.07% the flakes undertake a buckling instability.[38] Due to the competition between the buckling of the flakes and the adhesion with the substrate, a wavy pattern (ripples) appears on the flakes. Interestingly, the period of these ripples ($\lambda$) depends only on the mechanical properties of the flake and substrate:[38-40]

$$\lambda = h \left[\frac{8\pi^3(1-v_s^2)E_f}{3(1-v_f^2)E_s}\right]^{1/3}$$

where $h$ is the thickness of the flake, $v_s$ and $v_f$ correspond to the Poisson's ratio of substrate and flake and $E_s$ and $E_f$ are the Young's modulus of the substrate and flake, respectively. Therefore, for a known substrate one can extract the Young's modulus of InSe by measuring the period of ripples for flakes with different thickness. We have recently demonstrated the suitability of this method to probe the mechanical properties of 2D materials and we address the readership to Ref. [24] for more details on the technique.

Fig. 1a shows a cartoon of the fabrication process followed to apply the uniaxial compression to InSe flakes. A rectangular shape Gel-Film substrate is slightly bent to induce a uniaxial expansion of its topmost surface. Then InSe flakes are transferred on the bent surface by mechanical exfoliation with Nitto tape (see Materials and Methods) and the substrate stress is released yielding to the desired uniaxial compression. Fig. 1b shows a transmission mode optical microscopy image of an isolated InSe flake fabricated following this approach. The flake presents a marked wavy pattern because of the buckling instability induced ripples. The corresponding atomic force microscopy (AFM) topography image is shown in Fig. 1b with the thickness ranging from ~6L to ~24L (9.4 nm and 24 nm thick). The height extracted from the AFM dynamic mode measurements presented in this work contain a systematic offset that we determined to be 4.9 ± 0.5 nm in the case of InSe (this value should be subtracted from the AFM data in order to find the real flake thickness, see Section 2 of the Supplementary Information). This offset, which has been reported also for other 2D materials, is caused by the different interaction between the tip and either the substrate or the flake.[45] It is important to notice that this offset does not influence the determination of the Young's modulus of InSe and thus we did not subtract it from the data presented in this article. Fig. 1c shows the sinusoidal line profile shape of the ripples. From Fig. 1b-1c, one can see that the wavelength $\lambda$ of the ripples is strongly thickness dependent.

As seen in Equation (1), the Young's modulus can be determined by measuring the wavelength of the buckling induced ripples for flakes of different thickness. Fig. 2a shows six grayscale transmission mode optical images of the ripples produced on InSe flakes with different thicknesses. The wavelength of the ripples can be accurately extracted through the fast Fourier transform (FFT) of these images (Fig. 2b). Fig. 2c shows a summary of the wavelength values measured from 20 InSe flakes with thicknesses ranging from 6.4 nm to 32 nm (~1-2 layers to 40 layers). Note that rippled patterns with wavelengths smaller than ~1 µm and amplitude <10 nm cannot be well-resolved with optical microscopy.[24] The experimental data points follow a marked linear trend, as expected from Equation (1). From the slope $\lambda/h$ = 146 $\pm$ 11 we can determine the Young's modulus of InSe.Using the known values of the Poisson's ratio of PDMS (Gel-Film) $v_s$ = 0.5 and InSe flake $v_f$ = 0.27,[50, 51] and the measured Young's modulus of our Gel-Film $E_s$ = 492 $\pm$ 11 kPa (see the Supplementary Information of Ref. [24]),[24] we determine the Young's modulus of thin InSe flakes, $E$ = 23.1 ± 5.2 GPa, a value much smaller than the reported Young's modulus of other 2D materials.

In order to put this value in the more general context of 2D materials, we show in Fig. 3a a graphical comparison between the Young's modulus values for various isolated 2D materials available in the literature in a semi-logarithm scale plot. We indicate the highest and lowest Young's modulus reported in the literature through the error bar. The four differently colored regions correspond to semimetal, semiconductor, insulator and topological insulator groups, respectively. The value determined with buckling method of thin InSe Young's modulus is shown with a black dashed line and a surrounding blue shadow (that represents the uncertainty of our measurement result). According to the plot, the thin InSe flakes have a Young's modulus value which is around two orders of magnitude smaller than that of graphene and one order of magnitude lower than $MoS_2$ flakes. This value of the Young's modulus is among the lowest values reported for 2D materials up to date and is comparable only with that of metal-organic frameworks (MOFs, ~5 GPa).[53] Notably, this small value for the Young's modulus of InSe is in agreement with theoretical calculations, which predict an isotropic Young's modulus for monolayer InSe of 57 GPa and that this value is much smaller than the predicted values for $MoS_2$ and graphene.[50] All the details displayed in Fig. 3a are also summarized in Table 1 to facilitate a quantitative comparison between different materials.

The low Young's modulus value of InSe has implications in its applicability in flexible electronics, strain engineering or sensors. For example, applying or transferring strain to a 2D material with low Young's modulus requires less force than transferring the same amount of strain to a stiffer 2D material. To understand the role of the Young's modulus in the transfer of strain to a 2D materials deposited on a substrate, we performed a three-dimensional axisymmetric finite element analysis (FEA) using the software COMSOL Multiphysics (version 5.1). The results shown in Fig. 3b are calculated for a thin InSe with a thickness of 10 nm, diameter 20 μm and Young's modulus $E_{InSe}$ = 23.1 GPa, placed on a substrate with a thickness of 1000 μm and a diameter of 10000 μm. Additional details about the FEA calculation can be found in Section 3 of the Supplementary Information. The interface between the InSe flake and the substrate is modeled using perfect bonding. In each step of the simulation we let the substrate expand and we extract the total expansion induced in the InSe flake from which we calculate the strain transfer for all the different substrate Young's modulus values. We repeated the calculation for MoS$_2$ ($E_{MoS2}$ = 250 GPa) and graphene ($E_{Graphene}$ = 1 TPa).[5, 7, 24] The calculations show that, independently from the Young's modulus of the 2D material, for very small values of the substrate Young's modulus no strain is transferred from the substrate to the 2D flake, while for very large values the strain transfer approaches 100%. In between these two limits one can see that the strain transfer present in all three cases a similar sigmoidal shape and is shifted along the horizontal axis. The onset of transfer of each curve depends on the 2D material Young's modulus being the lowest for InSe and the largest for graphene. For 2D flakes with lower Young's modulus the strain transfer is larger than for flakes with larger modulus (given the same substrate Young's modulus). Since no atomistic details are taken into account in the simulation, the real absolute value of strain transfer for a given substrate Young's modulus can differ from the calculated one. Nevertheless, the general shape of the strain transfer curve and the trend observed should hold true for all the different 2D materials.

## Conclusions

In conclusion, we reported the experimental value of the Young's modulus of thin InSe flake using the buckling metrology method. We find an ultralow value in the range of 18 – 28 GPa, which makes thin InSe one of the most flexible two-dimensional materials. This low Young's modulus value can result highly valuable in applications such as force sensing (as the same force will produce a larger deformation in InSe than in other 2D semiconductors) or flexible electronic applications (were a small mismatch between the Young's modulus of the substrate and the 2D semiconductor is required to avoid stress accumulation after several flexing cycles). Also, we calculate that such superior flexibility may facilitate more effective strain transfer from a polymeric substrate to InSe flakes deposited on top, making InSe a very interesting material for applications in future straintronic devices.

## Materials and Methods

**Materials and sample fabrication.** The high-quality InSe crystal was grown by Bridgman method. With the atomic ratio In:Se = 52:48, high purity indium (99.9999%) and selenium (99.9999%) elements are sealed into a quartz crucible under the high vacuum conditions (2×10$^{-5}$ Pa). Then the quartz crucible was transferred into the rocking furnace and kept swing for 48 h at 650 ℃ to mix the melt completely. Then a vertical Bridgman furnace with a temperature gradient of 10 ℃/cm was used for crystal growth. Detailed characterization of the high-quality bulk materials can be found in Section 1 of the Supplementary Information. The elastomeric compliant substrate used in this work is commercially available polydimethylsiloxane (PDMS) supplied by Gel-Pak (WF 4x Gel-Film), whose Young's modulus ($E_s$ = 492 $\pm$ 11 kPa) has been determined in our previous work.[24] Mechanical exfoliation method with Scotch tape (3M®) and Nitto tape (Nitto Denko® SPV 224) was used for thin InSe flakes fabrication. Then the flakes were transferred onto a curved Gel-Film substrate to induce wrinkles. All the exfoliation processes are carried out in the ambient environment and at room temperature.

**Determination of wrinkle size and flake thickness.** The images of rippled InSe flakes onto PDMS were acquired with optical microscopy transmission mode (Motic® BA310 MET-T) in the case of wrinkles with wavelength larger than ~1 μm and with dynamic mode atomic force microscopy in the case of wavelengths smaller than 1 μm. Afterward, these rippled flakes were transferred onto SiO$_2$ (280 nm)/Si substrates for the AFM measurements. Gwyddion software was used for the quantitative analysis of the wavelength of the ripples on the flakes. The thickness of InSe flakes were determined by ezAFM from Nanomagnetics operating in dynamic mode. A Tap190Al-G by BudgetSensors with force constant 40 N/m and resonance frequency 300 kHz cantilever was used. To take into account the offset present in the thickness determination of 2D flakes by dynamic mode AFM,[45] one has to subtract a constant value of 4.9 nm from the raw AFM heights. See Section S2 of the Supplementary Information for more details on the estimation of this offset. Notice that all the thickness values given in the main text include this offset and thus to find the real thickness one has to subtract 4.9 nm from the reported height.

**Acknowledgements**

This project has received funding from the European Research Council (ERC) under the European Union's Horizon 2020 research and innovation program (grant agreement n° 755655, ERC-StG 2017 project 2D-TOPSENSE). EU Graphene Flagship funding (Grant Graphene Core 2, 785219) is acknowledged. RF acknowledges support from the Spanish Ministry of Economy, Industry and Competitiveness through a Juan de la Cierva-formación fellowship (2017 FJCI-2017-32919). QHZ acknowledges the grant from China Scholarship Council (CSC) under No. 201700290035. TW acknowledges support from the National Natural Science Foundation of China: 51672216.

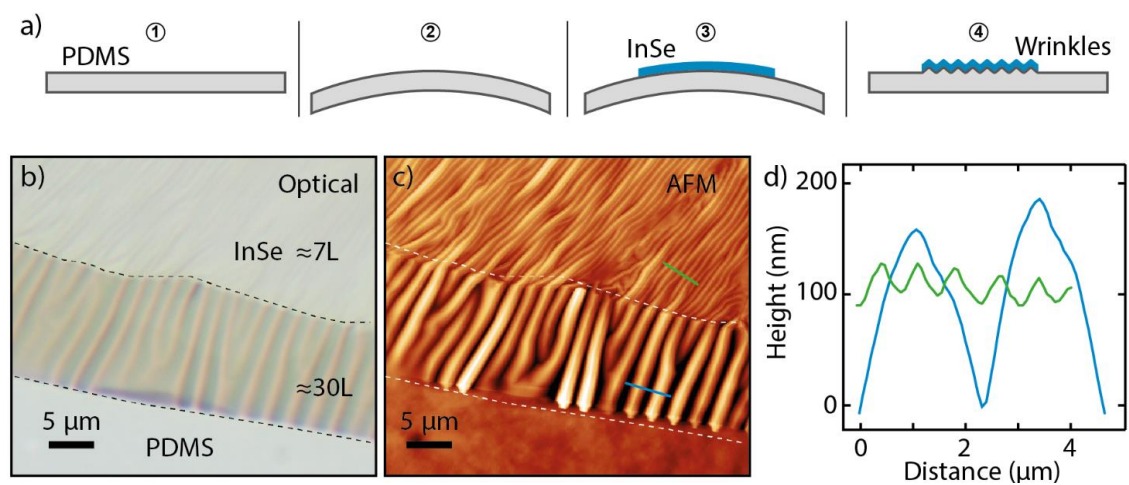

**Fig. 1** The wrinkled thin InSe flakes. a) The schematic diagram of fabricating wrinkled InSe flakes on Gel-Film substrate by using the buckling metrology method. b, c) Transmission mode optical pictures (b) and surface morphology recorded by AFM (c) of a wrinkled InSe flake with thicknesses (*h*) of ~6L and ~24L. d) The line profiles of the wrinkles with different thicknesses recorded at the positions marked in panel c (green line, 6L; blue line, 24L).

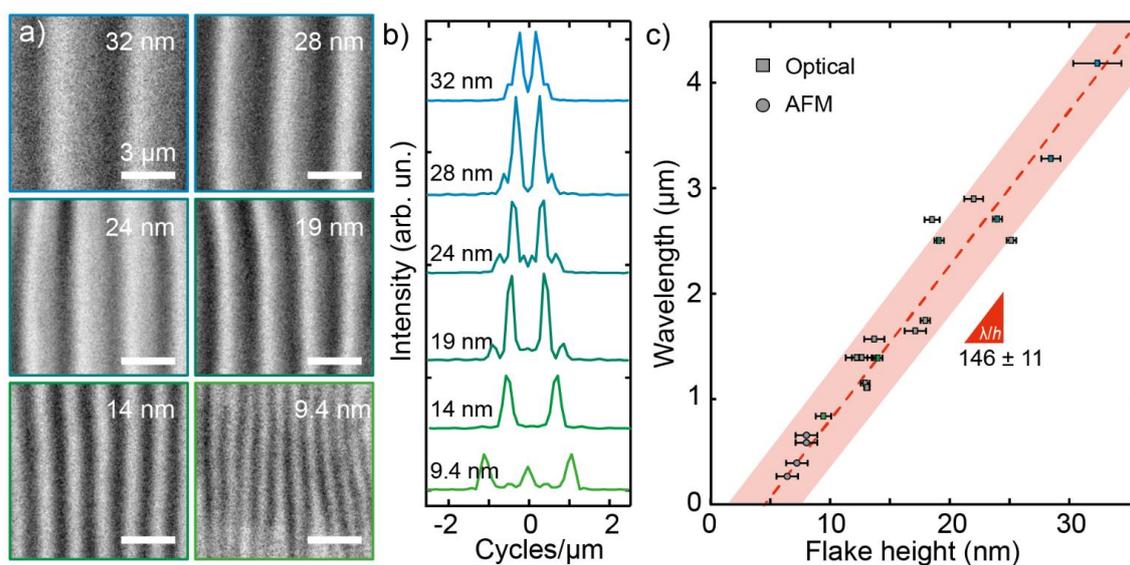

**Fig. 2** The determination of Young's modulus of thin InSe flakes. a) Grayscale transmission mode optical microscopy images of wrinkled pattern on InSe flakes with thicknesses going from 9.4 nm to 32 nm. b) The line cuts along the FFTs maxima extracted from the optical images in panel a. c) Relationship between the wavelength of the wrinkles and the thickness of the InSe flakes. The red dash line represents the linear fit based on the data and the light red shad area indicates the uncertainty of the fitting.

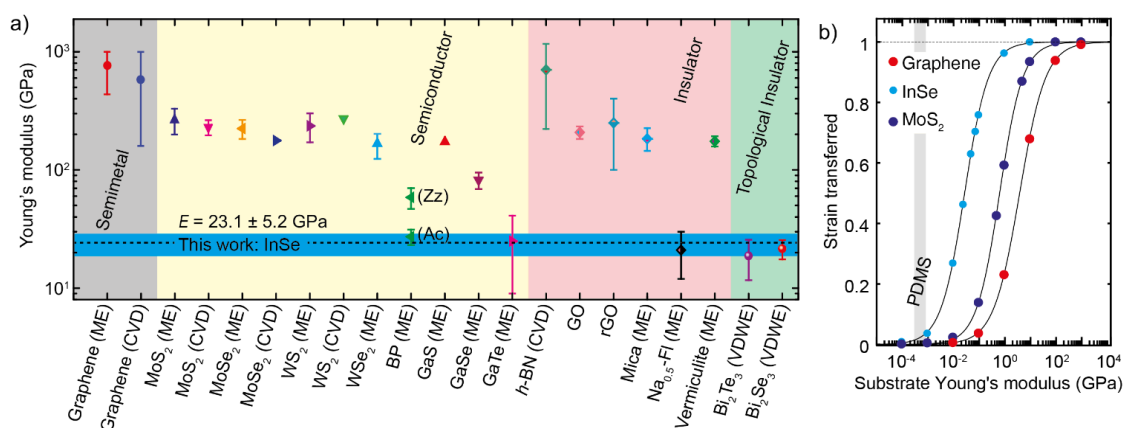

**Fig. 3** Summary of experimentally measured Young's modulus values of isolated two-dimensional species (semimetals, semiconductors, insulators and topological insulators). a) The different values of the Young's modulus of 2D materials reported in the literature are compared with the value of InSe obtained in this work $E = 23.1 \pm 5.2$ GPa (centered at the black dashed line with the blue shaded area indicating the experimental uncertainty). b) Strain transfer as a function of the substrate Young's modulus calculated from finite elements analysis for InSe, MoS$_2$ and graphene. The grey shaded regions indicate the Young's modulus values of polymer substrates typically used in flexible and printed electronics.

**Table 1** The comparison of Young's modulus of two-dimensional (2D) materials measured at room temperature and ambient environments. Legend: "Isolation method"; ME: mechanical exfoliation, CVD: chemical vapor deposition, VDWE: Van der Waals epitaxy. "Testing method"; (1) Spring constant scaling, (2) Nano-indentation, (3) Compliance maps, (4) Electrostatic deflection, (5) Blister test, (6) Constant force maps, (7) Nano-resonator, (8) Bimodal AFM, (9) buckling metrology method, (10) Micro-tensile method.

| Type | Materials | Isolation method | # of layers/ thickness | Young's modulus (GPa) (Testing method) | | Ref. |
|---|---|---|---|---|---|---|
| | | | | highest | lowest | |
| | InSe | ME | ~1-2L to 27L | 23.1±5.2 (9) | | This work |
| Semi-metal | Graphene | ME | 1-100L | 1000±100 (2) | 430 (1) | 5, 6 |
| | | CVD | 1L | 1000±50 (2) | 160 (2) | 7, 8 |
| semiconductor | $MoS_2$ | ME | 1-108L | 330±70 (2) | 200 (7) | 14, 15 |
| | | CVD | 1-2L | 264±18 (2) | 197±31 (5) | 21, 22 |
| | $MoSe_2$ | ME | 5-10L | 224±41 (9) | | 24 |
| | | CVD | 1-2L | 177.2±9.3 (10) | | 26 |
| | $WS_2$ | ME | 3-8L | 236±65 (9) | | 24 |
| | | CVD | 1L | 272±18 (2) | | 21 |
| | $WSe_2$ | ME | 4-12L | 167±7 (2) | 163±39 (9) | 24, 29 |
| | BP | ME | 15-25 nm | Zz:58.6±11.7, Ac: 27.2±4.1 (2) | | 32 |
| | GaS | ME | ~5-20 nm | 173±15 (2) | ~50 (2) | 36 |
| | GaSe | ME | ~10-30 nm | 82±13 (2) | | 36 |
| | GaTe | ME | ~10-30 nm | ~75 (2) | 25±16 (2) | 36 |
| Insulator | h-BN | CVD | 1 nm, 15 nm | 1160±100 (2) | 223±16 (2) | 41, 42 |
| | Graphene oxide | - | 1L | 207.6±23.4 (6) | | 43 |
| | | reduced | 1L | 250±150 (1) | | 44 |
| | Mica | ME | 2-14L | 200±30 (2) | 170±40 (1) | 46 |
| | $Na_{0.5}$ – Fluo | ME | 12-90L | 21±9 (1) | | 47 |
| | Vermiculite | ME | >2L | 175±16 (6) | | 48 |
| Topological insulator | $Bi_2Te_3$ | VDWE | 5-14 nm | 11.7-25.7 (2) | | 49 |
| | $Bi_2Se_3$ | VDWE | 7-12L | 17.48-25.45 (2) | | 52 |